# Femtocell Networks: A Survey

Vikram Chandrasekhar and Jeffrey G. Andrews, The University of Texas at Austin
Alan Gatherer, Texas Instruments

June 28, 2008
Contact: jandrews@ece.utexas.edu

## Abstract

The surest way to increase the system capacity of a wireless link is by getting the transmitter and receiver closer to each other, which creates the dual benefits of higher quality links and more spatial reuse. In a network with nomadic users, this inevitably involves deploying more infrastructure, typically in the form of microcells, hotspots, distributed antennas, or relays. A less expensive alternative is the recent concept of femtocells—also called home base-stations—which are data access points installed by home users to get better indoor voice and data coverage. In this article, we overview the technical and business arguments for femtocells, and describe the state-of-the-art on each front. We also describe the technical challenges facing femtocell networks, and give some preliminary ideas for how to overcome them.

## 1 Introduction

The demand for higher data rates in wireless networks is unrelenting, and has triggered the design and development of new data-minded cellular standards such as WiMAX (802.16e), 3GPP's High Speed Packet Access (HSPA) and LTE standards, and 3GPP2's EVDO and UMB standards. In parallel, Wi-Fi mesh networks also are being developed to provide nomadic high-rate data services in a more distributed fashion [1]. Although the Wi-Fi networks will not be able to support the same level of mobility and coverage as the cellular standards, to be competitive for home and office use, cellular data systems will need to provide service roughly comparable to that offered by Wi-Fi networks.

The growth in wireless capacity is exemplified by this observation from Martin Cooper of Arraycomm: "The wireless capacity has doubled every 30 months over the last 104 years". This translates into an approximately million-fold capacity increase since 1957. Breaking down these gains shows a 25x improvement from wider spectrum, a 5x improvement by dividing the spectrum into smaller slices, a 5x improvement by designing better modulation schemes, and a whopping 1600x gain through reduced cell sizes and transmit distance. The enormous gains reaped from smaller cell sizes arise from efficient spatial reuse of spectrum, or alternatively, a higher area spectral efficiency [2].

The main problem to this continued micro-ization of cellular networks is that the network infrastructure for doing so is expensive. A recent development are femtocells, also called home base-stations, which are short range, low cost and low power base-stations, installed by the consumer for better indoor voice and data reception. The user-installed device communicates with the cellular network over a broadband connection such as DSL, cable modem, or a separate RF backhaul channel. While conventional approaches require dual-mode handsets to deliver both in-home and mobile services, an in-home femtocell deployment promises fixed mobile convergence with *existing* handsets. Compared to other techniques for increasing system capacity, such as distributed antenna systems [3] and microcells [4], the key advantage of femtocells is that there is very little upfront cost to the service provider. Table 3 provides a detailed comparison of the key traits of these three approaches.

Studies on wireless usage show that more than 50% of all voice calls [Airvana, 5] and more than 70% of data traffic originates indoors [Picochip, 5]. Voice networks are engineered to tolerate low signal quality, since the required data rate for voice signals is very low, on the order of 10 kbps or less. Data networks, on the other hand, require much higher signal quality in order to provide the multi-Mbps data rates users have come to expect. For indoor devices, particularly at the higher carrier frequencies likely to be deployed in many wireless broadband systems, attenuation losses will make high signal quality and hence high data rates very difficult to achieve. This raises the obvious question: why not encourage the end-user to install a short-range low-power link in these locations? This is the essence of the win-win of the femtocell approach. The subscriber is happy with the higher data rates and reliability; the operator reduces the amount on traffic on their expensive macrocell network, and can focus its resources on truly mobile users. To summarize, the key arguments in favor of femtocells are the following.

**Better coverage and capacity**. Due to their short transmit-receive distance, femtocells can greatly lower transmit power, prolong handset battery life, and achieve a higher signal-to-interference-plus-noise ratio (SINR). These translate into improved reception—the so-called *five-bar* coverage—and higher capacity. Because of the reduced interference, more users can be packed into a given area in the same region of spectrum, thus increasing the area spectral efficiency [2], or equivalently, the total number of active users per Hz per unit area.

**Improved macrocell reliability**. If the traffic originating indoors can be absorbed into the femtocell networks over the IP backbone, the macrocell BS can redirect its resources towards providing better reception for mobile users.

**Cost benefits**. Femtocell deployments will reduce the operating and capital expenditure costs for operators. A typical urban macrocell costs upwards of $1K per month in site lease, and additional costs for electricity and backhaul. The macrocell network will be stressed by the operating expenses, especially when the subscriber growth does not match the increased demand for data traffic [Airvana, 5]. The deployment of femtocells will reduce the need for adding macro-BS towers. A recent study [6] shows that the operating expenses scale from $60K per year per macrocell to just $200 per year per femtocell.

**Reduced subscriber turnover**. Poor in-building coverage causes customer dissatisfaction, encouraging them to either switch operators or maintain a separate wired line whenever indoors. The enhanced home coverage provided by femtocells will reduce motivation for home users to switch carriers.

The goal of this article is to provide an overview for these end-user deployed infrastructure enhancements, and describe in more detail how the above improvements come about. We also describe the business and technical challenges that femtocells present, and provide some ideas about how to address them.

## 2  Technical Aspects of Femtocells

The capacity potential of femtocells can be verified rapidly from Shannon's law, which relates the wireless link capacity (in bits/second) in a bandwidth W Hz to the Signal-to-Interference plus Noise ratio (SINR). The SINR is a function of the transmission powers of the desired and interfering transmitters, path losses and shadowing during terrestrial propagation. Path losses cause the transmitted signal to decay as $Ad^{-\alpha}$, where A is a fixed loss, d is the distance between the transmitter and receiver, and α is the path-loss exponent. The key to increasing capacity is to enhance reception between intended transmitter-receiver pairs by minimizing d and α. Simultaneously, additional benefits in the network-wide spatial reuse can be obtained by—but not restricted to—exploiting diversity, and employing interference cancellation, interference suppression and interference avoidance techniques.

Femtocells enable a reduced transmit power, while maintaining good indoor coverage. Penetration losses insulate the femtocell from surrounding femtocell transmissions. Assuming a fixed receive power target with a path-loss propagation model (no fading), and denoting α (resp. β) as the outdoor (resp. indoor) path-loss exponent, overlaying an area $L^2$ with N femtocells results in a transmit power reduction of the order of $[10(α-β) \log_{10} L + 5 β \log_{10} N]$ dB. For example, choosing a cell dimension of L=1000 meters and N=50 femtocells, with equal path-loss exponents α=β=4, femtocells give a transmit power saving of nearly 34 dB. When the indoor path-loss exponent is smaller, say choosing β=2, the transmit power savings increase to nearly 77 dB.

To summarize, the capacity benefits of femtocells are attributed to:

1. Reduced distance between the femtocell and the user, which leads to a higher received signal strength.

2. Lowered transmit power, and mitigation of interference from neighboring macrocell and femtocell users due to outdoor propagation and penetration losses.

3. As femtocells serve only around 1-4 users, they can devote a larger portion of their resources (transmit power & bandwidth) to each subscriber. A macrocell, on the other hand, has a larger coverage area (500m-1 km radius), and a larger number of users; providing Quality of Service (QoS) for data users is more difficult.

Points 1 & 2 illustrate the dual improvements in capacity through increased signal strength and reduced interference. Point 3 shows that deploying femtocells will enable more efficient usage of precious power and frequency resources. The assumption here is that the wired broadband operator provides sufficient QoS over the backhaul. Otherwise backhaul capacity limitations could reduce the indoor capacity gains provided by femtocells.

**Example.** Consider a cellular OFDMA system with 100 active users. One scenario consists of a single macrocell serving all 100 users simultaneously, and the other scenario consists of 50 femtocells, with two active users in each femtocell. Figure 1 illustrates the cumulative distribution function of the normalized per user and sum throughput (normalized by the overall bandwidth), considering voice and data traffic. Simulations show a nearly 0.6 b/s/Hz gain in normalized median user throughput in femtocell deployments for voice only networks.

With data traffic, it is infeasible for a macrocell to provide data services to all 100 users simultaneously because of the limited transmission power and spectrum availability per user. We therefore consider a somewhat pathological case, in which the macrocell always schedules the 20 strongest users for transmission. On the other hand, femtocells can transmit simultaneously over the entire bandwidth. Compared to a macrocell, a femtocell deployment shows a normalized user throughput gain equaling 1.8 b/s/Hz and a huge system wide median sum throughput gain of nearly 250 b/s/Hz. This shows that the biggest benefits of femtocells are a massive improvement in the system spatial spectral efficiency.

## 3  Business Aspects of Femtocells

Even though femtocells offer savings in site lease, backhaul and electricity costs for the operator, they incur strategic investments. Operators will need to aggressively price femtocells despite tight budgets and high manufacturing costs, to compete with ubiquitous Wi-Fi. For example, the North American operator Sprint charges a subsidized price of $49.99 per *Airave* femtocell, for subscribing to a $30/month family plan. At the same time, the features that femtocells have to provide are in many ways more sophisticated than what is in a consumer grade Wi-Fi access point. The nascent femto vendors are facing cost targets set by the mature high-volume Wi-Fi market and by the demands of the operators for minimal subsidy to reduce Return-on-Investment (ROI) time. Consequently, *cost issues are in most cases, the central factor driving the selection of solutions to each technical challenge.*

Table 1 shows a predictive cost breakup in a femtocell network deployment, conducted by Airvana and Gartner [5]. On balance, it can be assumed that after 1.5 years, operator investment will be recovered, allowing future profits.

## 3.1 Current Standardization and Deployments

Given the aggressive cost challenges, standardization of requirements across customers is important to accomplish a low cost femtocell solution. Towards this end, a collaborative organization called the *Femto-Forum* comprised of operators and femtocell vendors was formed in 2007 with the objective of developing open standards for product interoperability.

Table 2 shows the current state of femtocell deployments. The North American operator Sprint, provides CDMA 1x EVDO services in Denver, Indianapolis and Tennessee; concurrently, a number of operators—Verizon and AT&T (USA), O2 Telefonica and Motorola (Europe) and Softbank (Japan)—are conducting femtocell trials prior to market. ABI Research [5] predicts 102 million users worldwide on more than 32 million femtocells by 2012.

## 4   Technical Challenges

This section overviews the key technical challenges facing femtocell networks.
1. Broadband Femtocells: Resource allocation, timing/synchronization and backhaul.
2. Voice Femtocells: Interference management in femtocells, allowing access to femtocells, handoffs, mobility and providing Emergency-911 services.
3. Network Infrastructure: Securely bridging the femtocell with the operator network over IP.

## 4.1 Physical and Medium Access layers: Broadband Femtocells

Confronting operators will be the dual problems of mitigating RF interference and efficiently allocating spectrum in femtocell networks. Interference mitigation will require innovative solutions since the low-cost target potentially necessitates scaled-down signal processing capabilities inside femtocells. The RF interference will arise from a) Macrocell to femtocell interference, b) Femtocell to femtocell interference and c) Femtocell to macrocell interference. The near-far effect—due to uneven distribution of receive power—is the main contributor for a) and c), while femtocell to femtocell interference is relatively smaller due to low transmit power and penetration losses.

**Challenge 1: How will a femtocell adapt to its surrounding environment and allocate spectrum in the presence of intra- and cross-tier interference?**

The 3GPP LTE and WiMAX standards ensure intra-cell orthogonality among macrocellular users and mitigate inter-cell interference through fractional frequency reuse. Since femtocells will be placed by end consumers, the ad-hoc locations of femtocells will render centralized frequency planning difficult.

Owing to the absence of coordination between the macrocell and femtocells and between femtocells, decentralized spectrum allocation between macrocell and femtocell users is an open research problem, which can provide answers to the following questions.

- Should macrocell and femtocell users be orthogonal through bandwidth splitting? Is there an "optimal" splitting policy? How does this vary with the femtocell density?
- Alternatively, with shared bandwidth (i.e. universal frequency reuse), what fraction of the spectrum should the macrocell and femtocells assign their users?
- Which of these two schemes is "better" in various configurations?

**Challenge 2: How will femtocells provide timing and synchronization?**

Femtocells will require synchronization to align received signals to minimize multi-access interference, and to ensure a tolerable carrier offset. Synchronization is also required so that macrocell users can handoff to a femtocell or vice versa, which is made more difficult due to absence of centralized coordination between them. With an IP backhaul, femtocells will experience difficulty in obtaining a time base that is immune to packet jitter. For 4G OFDMA air interfaces, ranging procedures to achieve timing (~ 1 μs) and frequency accuracy (~250 ppb) [7] [8] are needed for two reasons.

1. The inter-carrier interference arising from a carrier offset causes loss of sub-carrier orthogonality. Additionally, femtocells will have to compensate for frequency errors arising from the handset—which typically have poor oscillators.

2. In TDD systems, femtocells will require an accurate reference for coordinating the absolute phases for forward and reverse link transmissions and bounding the timing drift.

Although both points apply to the macrocell BS as well, the low cost burden and difficulty of synchronizing over backhaul will make efficient synchronization especially important for femtocells. Network solutions such as the IEEE-1588 Precision Timing Protocol over IP—with potential timing accuracy of 100 ns—and self-adaptive timing recovery protocols (e.g. the G.8261 standard) are promising. Another possibility is equipping femtocells with GPS for synchronizing with the macrocell, which relies on maintaining stable indoor satellite reception and keeping low costs in a price sensitive unit. Finally, high precision oven-controlled crystal oscillators may be used inside femtocells, incurring additional cost and periodic calibration.

**Challenge 3: How will backhaul provide acceptable QoS?**

IP backhaul needs QoS for delay sensitive traffic, and providing service parity with macrocells. Additionally, it should provide sufficient capacity, to avoid creating a traffic bottleneck. While existing macrocell networks provide latency guarantees within 15 ms, current backhaul networks are not equipped to provide delay resiliency. Lack of net neutrality poses a serious concern, except in cases where the wireline backhaul provider is the same company or in a tight strategic relationship with the cellular operator.

Another issue arises when femtocells usage occurs when the backhaul is already being used for delivering Wi-Fi traffic. Trials by Telefonica [5] reveal that when users employed Wi-Fi, the femtocells experienced difficulty transferring data and even low bandwidth services like voice. This is especially important, considering that improved voice coverage is expected to be a main driver for femtocells.

## 4.2 Physical and Medium Access layers: Voice Femtocells

For voice users, an operator faces two choices: either allocate different frequency bands to macrocell and femtocell users to eliminate cross-tier interference, or alternatively, serving both macrocell and femtocell users in the same region of bandwidth, to maximize area spectral efficiency. Considering the scarce availability of radio resources and ease of deployment, using the same region of bandwidth is preferable, if at all possible.

**Challenge 4: How will femtocells handle cross-tier interference?**

CDMA networks (without femtocells) employ fast power control to compensate for path-loss, shadowing and fading, and to provide uniform coverage. When femtocells are added, power control creates **dead-zones** (Figure 2) leading to non-uniform coverage. On the reverse link, a cell edge macrocell user transmitting at maximum power causes unacceptable interference to nearby femtocells. Consequently, cell edge femtocells experience significantly higher interference compared to interior femtocells. On the forward link, at the cell edge—where femtocells are most needed—macrocell users are disrupted by nearby femtocell transmissions, since they suffer higher path-loss compared to cell interior users.

> **Example**: Consider a CDMA reverse link with parameters:
> - Distance of macrocell user to macrocell= 500 meters
> - Femtocell radius= 20 meters
> - Distance of macrocell user to femtocell= 30 meters
> - Processing gain=128
> - Path-loss exponent= 4
> - Desired receive power= 0 dBm (1 mW)
>
> Interference Power from user at femtocell= $10*\log_{10}(500^4/30^4) = 48.87$ dB
> CDMA Interference suppression = $10*\log_{10}(128) = 21.07$ dB
> Post-processing Signal-to-Interference Ratio at femtocell= **-27.8 dB.**
> Reception is infeasible.

**Challenge 5: Should femtocells provide open or closed access?**

A closed access femtocell has a fixed set of subscribed home users—for privacy and security—that are licensed to use the femtocell. Open access femtocells, on the other hand, provide service to macrocell users, if they pass nearby. Radio interference is managed by allowing strong macrocell interferers to communicate with nearby femtocells.

Although open access reduces the macrocell load, the higher numbers of users communicating with each femtocell will strain the backhaul to provide sufficient capacity (related to Challenge 3) and raise privacy concerns for home users. Open access will need to avoid "starving" the paying home user—so they shouldn't ever see "all circuits busy." Since femtocells are typically marketed as offering flat-rate calling, open access will need to differentiate between the zero-tariff home users from the pay-per-minute visitor. For both reasons, operators are looking at hybrid models where some of the femto's resources are reserved for registered family members, while others are open for roamers.

**Challenge 6: How will handoff be performed in open access?**

In general, handoff from a femtocell to the macrocell network is significantly easier—as there is only one macro BS—as compared to handoffs from the macrocell to the femtocell. Current 2G and 3G systems broadcast a neighbor list which a mobile attached to the current cell uses to learn where to search for potential handover cells. Such a handoff protocol does not scale to the large numbers of femtocells that "neighbor" (actually underlay) the macrocell, and the underlying network equipment isn't designed to rapidly change the lists as femtocells come and go. This motivates 4G handover procedures to take the presence of femtocells into consideration.

In open access, channel fluctuations may cause a passing macrocell user to perform multiple handovers. In co-channel deployments, Claussen *et al* [10] have proposed auto configuration that periodically reduces the pilot power inside a femtocell when no active calls are in progress thereby minimizing handoffs from passing macrocell users. An open research area is to develop low complexity algorithms for predicting the dwell time before handing off a macrocell user onto a nearby femtocell. Yeung and Nanda [11] have proposed controlling handoff events by choosing velocity thresholds based on the user mobility and sojourn times when a macrocell user travels in the vicinity of a femtocell.

**Challenge 7: Can subscribers carry their femtocells for use outside the home area?**

The portability of the femtocell presents a conundrum: Unlike Wi-Fi networks that operate in unlicensed spectrum in which radio interference is not actively managed, femtocell networks will operate in licensed spectrum. Femtocell mobility can cause problems when a subscriber with operator *A* carries their femtocell to another location where the only service provider is a rival operator *B*. In such a scenario, should the femtocell be allowed to transmit on operator *B*'s spectrum? Viable options are providing GPS inside femtocells for location tracking and locking the femtocell within a geographical area. Alternatively, inter-operator agreements facilitate charging the home subscriber on roaming.

**Challenge 8: How will femtocells provide location tracking for Emergency-911 and should they service nearby macrocell users with poor coverage?**

Government mandated Emergency-911 services require operators to provision femtocells for transmitting location information during emergency calls. Femtocell location may be obtained by either using GPS inside femtocells (added cost with possibly poor indoor coverage), or querying the service provider for location over the backhaul, or gathering information from the macrocell providing the femtocell falls within the macrocell radio range, or even inferring the location from the mobile position (estimated by the macro network) at handoff to the femtocell.

Ethical/legal dilemmas can arise on whether a femtocell should service macrocell users with poor outdoor coverage for making emergency calls, if they are located within its radio range. In open access networks, this problem can be solved by handoff. Closed access femtocells should be provisioned to allow communication with unsubscribed users in the event of emergencies.

## 4.3 Network Infrastructure

In a femtocell environment, the operator will need to provide a secure and scalable interface for the femtocell over IP, at a reasonable cost. Traditional Radio Network Controllers (RNCs) are equipped to handle tens to hundreds of macrocells. *How will they provide equal parity service to femtocells over the internet?*

Three network interfaces have been proposed, of which the IMS/SIP and UMA based interfaces appear to be the architectures of choice.

**Iu-b over IP**: Existing RNCs connect to femtocells through standard Iu-CS (circuit-switched) and Iu-PS (packet-switched) interfaces present in macrocell networks. The advantage is that the Capex is comparatively low insofar as the operator can leverage existing RNCs. The shortcomings are the lack of scalability, and that the interface is not yet standardized.

**IMS/SIP:** The Internet Media Sub-System/Session Initiation Protocol interface provides a core network residing between the femtocell and the operator. The IMS interface converts subscriber traffic into IP packets and employs Voice over IP (VoIP) using the SIP protocol, and coexists with the macrocell network. The main advantages are scalability and rapid standardization.

Disadvantages include the Capex for upgrade, and Opex in maintaining two separate core networks for the macrocell and femtocell respectively.

**RAN gateway based UMA:** A Radio Access Network (RAN) gateway exists between the IP network and the operator network, aggregating traffic from femtocells. This gateway is connected to the operator network using a standard Iu-PS/CS interface. Between the femtocell and the RAN gateway, the UMA (Unlicensed Mobile Access) protocol makes use of secure IP tunneling for transporting the femtocell signals over the internet. Current UMA-enabled services such as T-Mobile's *Hotspot@Home* require dual-mode handsets for switching between in-home Wi-Fi and outdoor cellular access. Integrating the UMA client inside femtocells, rather than the mobile, would enable future deployments support use of legacy handsets.

## 5   Research Directions

We now consider the key areas and tools for conducting further femtocell research, with the goal of designing efficient femtocell architectures.

### 5.1 Interference Management

Owing to the ad-hoc topology of femtocell locations, interference suppression techniques alone will prove ineffective in femtocell networks. Successive Interference Cancellation—in which each user subtracts out the strongest neighboring interferers from their received signal—appears promising initially, but cancellation errors quickly degrade its usefulness [12]. Consequently, an interference avoidance approach wherein users *avoid* rather than *suppress* mutual interference is more likely to work well in geography-dependent femtocell networks. The low cost requirement is likely to influence the design of low complexity femtocell BS receivers—simple matched filter processing for example—and low complexity transmission schemes for sensing nearby available frequency channels to avoid collisions.

In CDMA femtocell networks with universal frequency reuse, for example, interference avoidance—through time-hopping and directional antennas—provides a 7x improvement in system capacity [9], when macrocell and femtocell users share a common bandwidth.

**Frequency- and Time-hopping.** In GSM networks, the interference avoidance offered through *slow frequency hopping* enables femtocell users and nearby transmitting macrocell users to avoid consistent mutual interference. Similarly, frequency-hopped OFDMA networks can use random sub-channel assignments in order to decrease the probability of persistent collision with neighboring femtocells.

In time-hopped CDMA, the CDMA duration $G\,T_c$ ($G$ is the processing gain and $T_c$ is the chip period) is divided into $N_{hop}$ hopping slots, where each user randomly selects a hopping slot for transmission and remains silent during the remaining slots. Random time-hopping reduces the average number of interfering users by a factor of $N_{hop}$, while trading-off the processing gain. The tradeoff is that femtocells are accommodated by changing the way an existing CDMA macrocell network operates.

**Directional Antennas** inside femtocells offer interference avoidance, with zero protocol overhead, by restricting radio interference within an antenna sector. Providing a reasonable unit cost and easy end user deployment are the key challenges confronting this approach.

**Adaptive Power Control** strategies vary the femtocell receive power target depending on its location. Commercial femtocells such as Sprint's *Airave* femtocell tackle cross-tier interference using an "automatic adaptation" protocol that adjusts the femtocell transmit power. Over the forward link in closed access, Ericsson [13] has proposed reducing interference to macrocell users by reducing the femtocell transmit power with increasing distance from the macro BS. The tradeoff is the decreased home coverage at far-off femtocells. Over the reverse link in closed access, we suggest providing a higher receive power target to femtocell users relative to macrocell users [9], which will vary based on femtocell location.

Figure 3 shows the reductions in the outage probability for a femtocell user (the femto-macro receive power ratios are 1 and 10), in conjunction with interference avoidance using CDMA time hopping and antenna sectoring. For open access, Kishore *et al.* [14] propose overcoming the near-far effect by extending femtocell coverage—allow a user to communicate to a macrocell only if their channel gain is appreciably higher—at the expense of increased interference between neighboring femtocells. Multi-mode phones [7] switch protocols in closed-access systems for mitigating interference, for example, transmit in WCDMA mode inside a femtocell, and revert to GSM mode otherwise.

## 5.2 MIMO Femtocells

Multiple antennas at the transmitter and/or the receiver (MIMO) exploit the spatial diversity of the wireless channel. Femtocells can perform temporal link adaptation through adaptive modulation and coding; additionally, MIMO spatial link adaptation will enable a femtocell to switch between providing high data rates and robust transmission. High data rates are obtainable by transmitting multiple spatial streams (spatial multiplexing) over high SINR links. Over low SINR links, MIMO provides robustness through open and closed loop diversity schemes such as space-time codes and beam forming. Interesting areas for future research are a) link adaptive mode switching for femtocells between diversity and spatial multiplexing [15], b) analyzing the effect of channel state information errors induced by co-channel interference on MIMO femtocell performance, c) the complexity limitations of MIMO femtocell receivers, which may be significant vs. macrocell receivers due to cost considerations, and d) channel models for MIMO femtocells, since the diversity characteristics may be very different from macrocells.

## 6 Conclusions

Femtocells have the potential to provide high quality network access to indoor users at low cost, while simultaneously reducing the burden on the whole system. This article has identified the key benefits of femtocells, the technological and business challenges, and research opportunities. From a technical standpoint, operators face challenges in providing a low cost solution, while mitigating RF interference, providing QoS over the IP backhaul, and maintaining scalability. From a business perspective, generating long-term revenue growth and overcoming initial end user subsidies are key challenges.

## 7 Acknowledgements

The authors would like to thank the reviewers for providing excellent feedback, which greatly improved the quality and readability of this article. The authors also thank Andrew Hunter (UT) and Shalinee Kishore (Lehigh) for their comments and suggestions on the paper, and Vijay Sundararajan (Texas Instruments) for ongoing feedback on our work.

# 8  References

[1]     Tropos Networks, "Picocell Mesh: Bringing Low-Cost Coverage, Capacity and Symmetry to Mobile WiMAX," White Paper.

[2]     M.–S Alouini and A.J. Goldsmith, "Area Spectral Efficiency of Cellular Mobile Radio Systems," *IEEE Transactions on Vehicular Technology*, vol. 48, no. 4, pp. 1047 – 1066, July 1999.

[3]     A. Saleh, A. Rustako and R. Roman, "Distributed Antennas for Indoor Radio Communications," *IEEE Transactions on Communications*, vol. 35, no. 12, pp. 1245 – 1251, Dec. 1987.

[4]     Chih-Lin I, L.J. Greenstein and R.D. Gitlin, "A Microcell/Macrocell Cellular Architecture for Low- and High-Mobility Wireless Users," *IEEE Journal on Selected Areas on Communication*, vol. 11, no. 6, pp. 885 – 891, Aug. 1993.

[5]     Presentations by ABI Research, Picochip, Airvana, IP.access, Gartner, Telefonica Espana, *2$^{nd}$ International Conference on Home Access Points and Femtocells*, [Online]. Available: http://www.avrenevents.com/dallasfemto2007/purchase_presentations.htm

[6]     Analysys, "Picocells and Femtocells: Will indoor base-stations transform the telecoms industry?," [Online]. Available: http://research.analysys.com

[7]     Picochip, "The Case For Home Base Stations," White Paper, April 2007.

[8]     J. G. Andrews, A. Ghosh, and R. Muhamed, *Fundamentals of WiMAX.* Prentice-Hall, 2007.

[9]     V.Chandrasekhar and J.Andrews, "Uplink Capacity and Interference Avoidance in Two-Tier Femtocell Networks," to appear, *IEEE Trans. On Wireless Communications*, Available at http://arxiv.org/abs/cs.NI/0702132

[10]    Lester T.W. Ho and Holger Claussen, "Effects of User-Deployed, Co-Channel Femtocells on the Call Drop Probability in a Residential Scenario," *The 18$^{th}$ Annual IEEE International Symposium on Personal, Indoor and Mobile Communications*, pp.1 – 5, Sep 2007.

[11]    Kwan L. Yeung and Sanjiv Nanda, "Channel Management in Microcell/Macrocell Cellular Radio Systems," *IEEE Trans. On Vehicular Technology,* vol. 45, no. 4,  pp. 601 – 612, Nov 2006.

[12]    S. Weber, J. G. Andrews, X. Yang, and G. de Veciana, "Transmission Capacity of Ad Hoc Networks with Successive Interference Cancellation," *IEEE Trans. on Information Theory*, vol. 53, no. 8, pp. 2799 – 2814, Aug. 2007.

[13] Ericsson, "Home NodeB Output Power," *3GPP TSG Working Group 4 meeting*, [Online]. Available: http://www.3gpp.org/ftp/tsg_ran/WG4_Radio/TSGR4_43bis/Docs/

[14] S. Kishore, L.J. Greenstein, H.V. Poor, and S.C. Schwartz, "Uplink User Capacity in a Multicell CDMA System with Hotspot Microcells," *IEEE Trans. On Wireless Communications,* vol. 5, no. 6, pp. 1333 – 1342, June 2006.

[15] A. Forenza, M.R. McKay, A. Pandharipande, R.W. Heath, and I.B. Collings, "Adaptive MIMO transmission for Exploiting the Capacity of Spatially Correlated Channels," *IEEE Transactions on Vehicular Technology*, vol. 56, no. 2, pp. 619 – 630, Mar. 2007.

## Biographies

Vikram Chandrasekhar is a Ph.D. candidate at UT Austin. He completed his B.S. at IIT Kharagpur and his M.S. at Rice University. His current research focuses on fundamental limits and algorithms for microcellular and hotspot-aided broadband cellular networks. He has held industry positions at National Instruments, Freescale, and Texas Instruments.

Jeffrey G. Andrews received a B.S. from Harvey Mudd College and an M.S. and Ph.D. in electrical engineering from Stanford University. He is an Associate Professor (effective 9/08) in the ECE Department at UT Austin, where he is the Director of the Wireless Networking and Communications Group. He is author of *Fundamentals of WiMAX* (Prentice-Hall 2007) and recipient of the 2007 NSF CAREER award. He has held industry positions at Intel and Qualcomm.

Alan Gatherer received a B.Eng. from Strathclyde University, and an M.S. and Ph.D. in electrical engineering from Stanford University. He is the CTO for the Communications Infrastructure and Voice Business and a Distinguished Member of Technical Staff at Texas Instruments. He is responsible for the strategic development of TI's digital baseband modems for 3G and 4G wireless infrastructure as well as high performance medical products.

# Figure 1: Femtocell vs. Macrocell Throughput

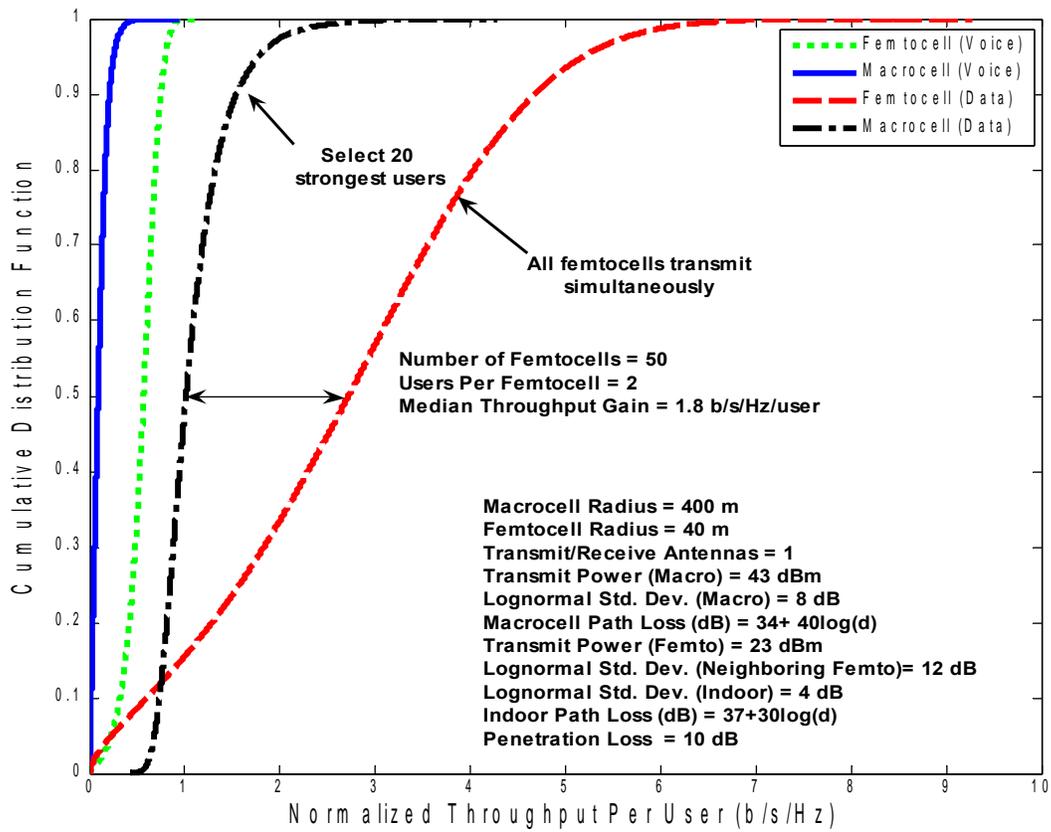

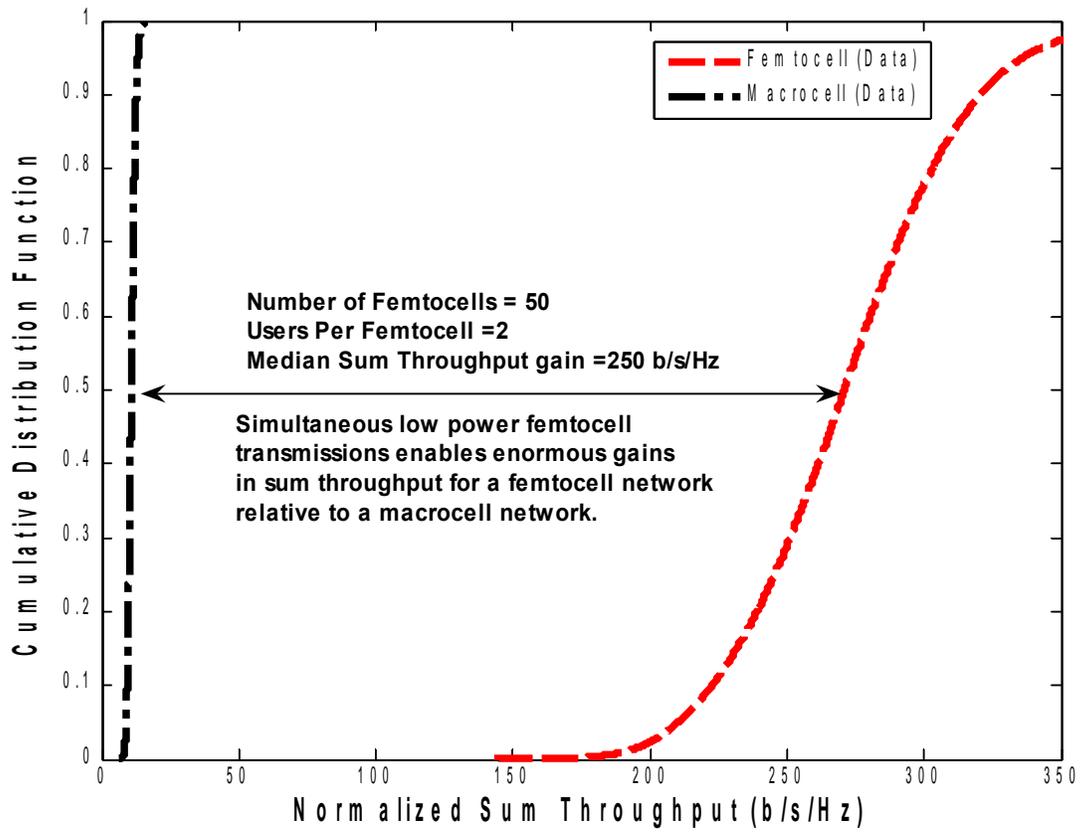

**Figure 2: Dead zones in CDMA femtocell networks**

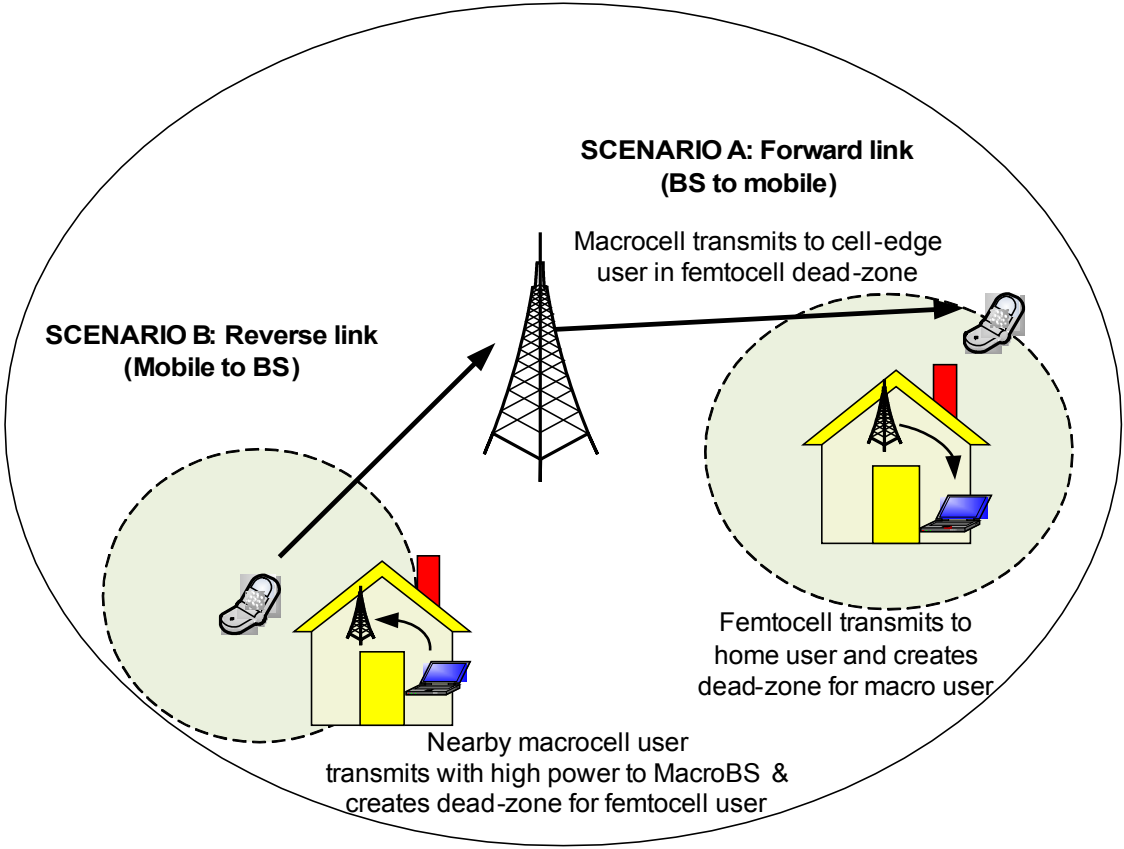

**Figure 3: Outage probability in a closed-access femtocell**

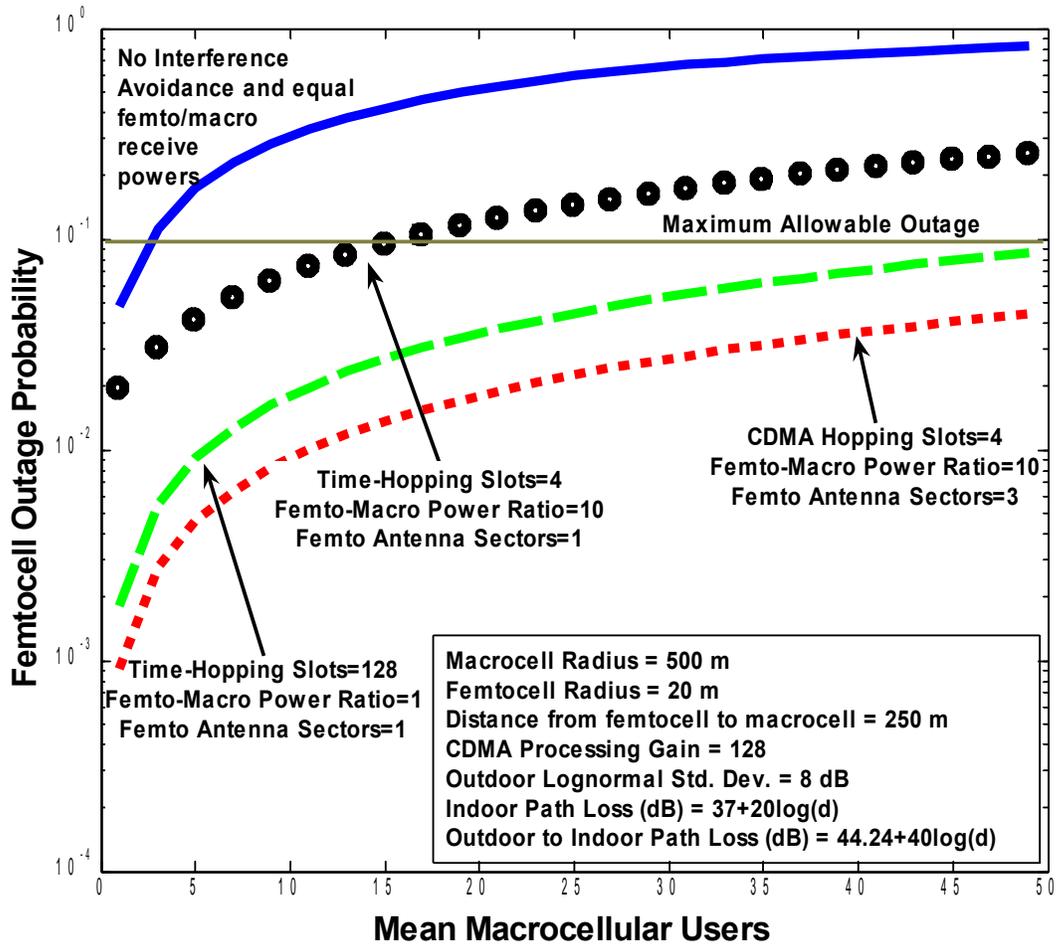

# Table 1: Predictive Return on Investment from Femtocell Deployments

| | |
|---|---|
| Airvana | **New Core network infrastructure**<br>Capital Expenditure = $35 per consumer<br><br>**Subscriber Addition**<br><br>a) Femtocell Price = $200<br>b) Price to Consumer = $50<br>c) Marketing Expense = $50<br>Net Expense per additional subscriber = $200<br><br>⎫ Fixed expense per consumer = $235<br><br>**Monthly Revenues**<br><br>a) Customer Revenue = $25<br>b) Network Operating Expenditure = $8<br>c) Customer Support = $3<br><br>⎫ Revenue per consumer = $14 per month<br><br>⎫ Implied ROI equals 16 months |
| Gartner | **Monthly Revenue**<br><br>1) Electricity savings from reduced macro BS power consumption = $0.50 per consumer<br><br>2) Monthly savings from reduction in number of macrocells = $1.50 per consumer<br><br>3) Monthly Revenue from indoor data downloads = $5 per consumer<br><br>4) Bundled voice services, broadband services over LTE/WiMAX/UMB = $2.50 per consumer<br><br>⎫ Revenue Per consumer = $9.50 per month<br><br>**Monthly Expenses**<br><br>1) Free voice calls inside femtocell = $2.50 per consumer<br><br>2) Operating expenditure for core network upgrades = $2 per consumer<br><br>⎫ Expense Per consumer = $4.50 per month<br><br>⎫ Implied ROI with $100 subsidy per consumer equals 20 months |

## Table 2: Existing Femtocell Offerings

| Manufacturer | Partner/Operator | Region | Technology |
|---|---|---|---|
| Samsung (Ubicell) | Sprint (Airave) | North America | a) IS-95, CDMA2000 1xEV-DO b) WCDMA |
| AirWalk Communications | | North America | CDMA 1x RTT & 1x-EVDO |
| Ericsson | | Europe | GSM/3GPP UMTS |
| Airvana | Nokia-Siemens | | 3GPP UMTS |
| Alcatel-Lucent | | North America | 3GPP UMTS |
| Axiom Wireless | PicoChip | United Kingdom | a) 3GPP UMTS b) WiMax |
| IP Access (Oyster) | PicoChip | United Kingdom | 3GPP UMTS |
| Ubiquisys (Zonegate) | Kineto wireless, Google | United Kingdom | 3GPP UMTS/HSPA |

## Table 3: Femtocells, Distributed Antennas and Microcells

| Infrastructure | Expenses | Features |
|---|---|---|
| **Femtocell:** Consumer installed wireless data access point inside homes, which backhauls data through a broadband gateway (DSL/Cable/Ethernet/WiMAX) over the internet to the cellular operator network. 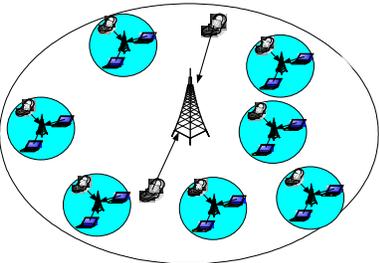 | **Capex.** Subsidized femtocell hardware. **Opex.** a) Providing a scalable architecture to transport data over IP, b) Upgrading femtocells to newer standards. | **Benefits.** a) Lower cost, better coverage and prolonged handset battery life from shrinking cell-size, b) Capacity gain from higher SINR and dedicated BS to home subscribers and c) Reduced subscriber churn **Shortcomings.** a) Interference from nearby macrocell and femtocell transmissions limits capacity and b) Increased strain on backhaul from data traffic may affect throughput. |

| **Distributed Antennas:** Operator installed spatially separated antenna elements (AEs) connected to a Macro BS via a dedicated fiber/microwave backhaul link. 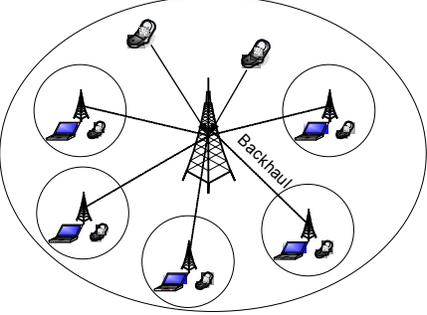 | **Capex.** AE & backhaul installation.<br><br>**Opex.** AE maintenance and backhaul connection. | **Benefits.** a) Better coverage since user talks to nearby AE, b) Capacity gain by exploiting both macro- and micro-diversity (using multiple AEs per macrocell user).<br><br>**Shortcomings.** a) Does not solve the indoor coverage problem, b) RF interference in the same bandwidth from nearby AEs will diminish capacity and c) Backhaul deployment costs may be considerable. |
|---|---|---|
| **Microcells:** Operator installed cell-towers, which improve coverage in urban areas with poor reception. 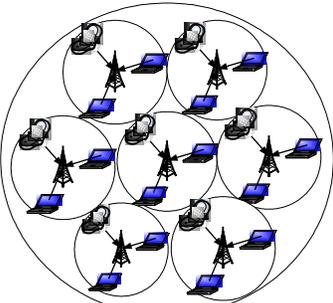 | **Capex.** Installing new cell-towers.<br><br>**Opex.** Electricity, site lease and backhaul. | **Benefits.** a) System capacity gain from smaller cell size, b) Complete operator control.<br><br>**Shortcomings.** a) Installation and maintenance of cell-towers is prohibitively expensive b) Does not completely solve indoor coverage problem. |